# On the definition of chirality and enantioselective fields

Jun-ichiro Kishine,[a] Hiroaki Kusunose,[b] and Hiroshi M. Yamamoto*[c]

[a]  Jun-ichiro Kishine
     Department of Natural Sciences,
     The Open University of Japan,
     Wakaba 2-11, Chiba261-8586, Japan

[b]  Hiroaki Kusunose
     Department of Physics, Meiji University
     Higashi-Mita, Taka-Ku, Kawasaki 214-8571 Japan

[c]  Hiroshi M. Yamamoto
     Research Centre of Integrated Molecular Systems,
     Institute for Molecular Science
     Myodaiji, Okazaki 444-8585 Japan
     e-mail: yhiroshi@ims.ac.jp

**Abstract**

In solid state physics, any symmetry breaking is known to be associated with emergence of an order parameter. However, the order parameter for molecular and crystal chirality, which is a consequence of parity and mirror symmetry breaking, has not been known since its discovery. In this article, the authors show that the order parameter for chirality can be defined by electric toroidal monopole $G_0$. By this definition, one becomes able to discuss external filed that can distinguish two different enantiomers only by physical fields. In addition, dynamics and fluctuations of the order parameter $G_0$ can be discussed, with which one can obtain fruitful insights on a spin filtering effect called CISS (Chirality Induced Spin Selectivity). Emergence of time-reversal-odd dipole $M_z$ by time propagation of $G_0$ quantities is discussed to explain the enantioselective effect (chiral resolution) at a ferromagnetic surface.

1. **Introduction: historical background**

CISS (Chirality Induced Spin Selectivity), an electron's spin-filtering effect by a chiral molecule [1-5], is believed to be a quantum mechanical phenomenon which only takes place in chiral molecules and crystals. However, quantum mechanical treatment of chiral media is not so trivial as one might think for the first thought because it is not a structureless point-like concept. Therefore, in order to establish the starting point for unveiling microscopic mechanism of CISS, it is pivotal to explicitly give the definition of the chirality in terms of quantum mechanical language.

Historically, in early 19th century, Arago and then Biot discovered the phenomena of natural optical activity (NOA). Later, in 1848, Pasteur ascertained that the NOA is a direct consequence of molecular or crystal asymmetry. In 1894, Lord Kelvin, in his Boyle Lecture [6], coined the term chirality with the definition, "any geometrical figure, or group of points, is chiral, if its image in a plane mirror, ideally realized, cannot be brought to coincide with itself." After the discovery of quantum mechanics, the existence of chiral molecules has puzzled many scientists because only one-handed enantiomeric state cannot be an eigenstate of the Hamiltonian [7, 8]. An important point here is that two enantiomeric states, both of which are required to constitute the complete eigenstates, are separated by double-well potential and the molecule is allowed to stay at one of these two states in a finite period of time because of symmetry breaking. (The time scale depends on the height of potential barrier and system size.) Barron has proposed to extend the definition of chirality from such a structural chirality to dynamics of materials and fields in his discussions on true/false chirality [9-11]. The most important aim of his discussion was to describe the necessary conditions for absolute asymmetric synthesis, which might be a cause of homochirality in nature. However, such a discussion becomes heavily complicated when non-equilibrium states are considered, which precludes universal understanding of physical fields that can recognize structural chirality.

In this short review, we try to describe the details of chiral symmetry breaking in terms of quantum mechanics with a knowledge of solid state physics, because the symmetry breaking is one of the central concepts for this field so that the methodology is well established. Through this method, we redefine the chirality in terms of the rank-zero multipole (monopole) basis. This new definition leads to unambiguous notion of structural chirality as well as enantio-selective fields. Following this argument, we also discuss recent proposals on the microscopic mechanism of the CISS phenomena.

## 2. Order parameter for chiral object

As has been introduced in the previous section, chirality refers to an existence of two objects, called enantiomers, which are interconnected by mirror operation but cannot overlap one another. This definition of chirality is empirically natural, but it is ambiguous for chiral objects in discrete point groups at quantum mechanical level (here, we use the term "chiral objects" to refer to molecules and crystals possessing chirality). In order to establish a concrete description for such microscopic chiral objects, we then develop a proper representation scheme based on the concept of electronic multipoles. The concrete description of chirality in an appropriate basis will also lead to the discussion of physical fields in the next section that can recognize the handedness of chiral objects, which should be called "enantioselective fields".

In the continuous SO(3) rotational group with inversion operation, parity (inversion) symmetry breaking is identical to mirror symmetry breaking. The situation becomes complicated when point groups, or subgroups of SO(3), are considered in order to express chirality of fixed molecules and crystals. In such a case, the inversion and mirror symmetries are independent, and even NOA can no more be as unique signature of chirality. It is important, however, to point out that none of the inversion or mirror symmetries exist for chiral point groups. Namely, the chiral point groups consist of purely rotational symmetry operations. At the same time, the difference between polar, weak and strong gyrotropic (WG and SG), and chiral point groups appear within non-centrosymmetric point groups having no inversion symmetry operation (Table 1).

The difference between chiral and other point groups must reflect in the character of their wave function and associated electronic degrees of freedom. To extract the specific character of the wave function, it is useful to express any electronic degrees of freedom in terms of the complete symmetry-adapted (multipole) basis set. For instance, the presence of the electric dipole moment discriminates a polar wave function from non-polar ones. Such a description can be generalized to any point groups, in which a proper multipole moment can distinguish different symmetry of wave functions as shown in Fig. 1. The presence of electric dipole moment is a manifestation of the *s-p* hybrid orbital, $\phi_s + \phi_{px}$ (Fig 1a). Similarly, *p-d* hybrid orbital, $\phi_{pz} + \phi_{dx^2-y^2}$ is characterized by the electric toroidal quadrupole (Fig 1b). On the other hand, the time-reversal-symmetry (TRS) broken states, $\phi_s + i\phi_{px}$ and $\phi_{pz} + i\phi_{dx^2-y^2}$ carrying orbital angular momenta (dipole and quadrupole), are characterized by the finite magnetic toroidal dipole and magnetic quadrupole moments, respectively (Fig 1c and 1d).

In this way, any wave functions under particular point group *with or without* TRS breaking are represented by the active moment of four type of electronic multipoles, *i.e.*, electric ($Q$), magnetic ($M$), electric toroidal ($G$), and magnetic toroidal ($T$) multipoles, whose properties are summarized in Table 2. The electric and magnetic multipoles are familiar in classical electromagnetism, which are polar (true) and axial (pseudo) tensors with time-reversal ($\mathcal{T}$)-even and $\mathcal{T}$-odd symmetries, respectively. On the other hand, less familiar electric toroidal and magnetic toroidal multipoles are axial (pseudo) and polar (true) tensors, which are counterparts of the ordinary electric and magnetic multipoles in the parity sector. Each multipole has its rank $l$, *e.g.*, monopole is $l=0$, dipole is $l=1$, quadrupole is $l=2$, and so on, and the polar (true) tensor is characterized by the parity $(-1)^l$, while the axial (pseudo) tensor by $(-1)^{l+1}$ against the inversion operation. Therefore, the even-rank polar and odd-rank axial tensors are parity ($\mathcal{P}$) even, while others are $\mathcal{P}$-odd. It is important to note that these four multipoles constitute a complete basis set, and hence they can describe any of symmetry-adapted electronic states. For further detail, refer to the literatures, [12-14].

Table 1. Active multipoles belonging to the totally symmetry irreducible representation for non-centrosymmetric point groups except for C$_1$. The abbreviations, P, C, SG, and WG denote Polar, Chiral, Strong Gyrotropic (NOA Active), and Weak Gyrotropic (NOA Inactive) point groups, respectively. The shaded rows represent chiral point groups. Full version of this table can be found in Table XVI in ref. [13].

| Crystal System | Point Group | P | C | SG | WG | Active Multipole Moment | | |
|---|---|---|---|---|---|---|---|---|
| | | | | | | Monopole | Dipole | Quadrupole |
| Cubic | O | | ✔ | ✔ | | $G_0$ | | |
| | T$_d$ | | | | | | | |
| | T | | ✔ | ✔ | | $G_0$ | | |
| Tetragonal | D$_4$ | | ✔ | ✔ | | $G_0$ | | $G_u$ |
| | D$_{2d}$ | | | ✔ | | | | $G_v$ |
| | C$_{4v}$ | ✔ | | | ✔ | | $Q_z$ | |
| | C$_4$ | ✔ | ✔ | ✓ | | $G_0$ | $Q_z$ | $G_u$ |
| | S$_4$ | | | ✔ | | | | $G_v, G_{xy}$ |
| Orthorhombic | D$_2$ | | ✔ | ✔ | | $G_0$ | | $G_u, G_v$ |
| | C$_{2v}$ | ✔ | | ✓ | | | $Q_z$ | $G_{xy}$ |
| Monoclinic | C$_2$ | ✔ | ✔ | ✔ | | $G_0$ | $Q_z$ | $G_u, G_v, G_{xy}$ |
| | C$_s$ | ✔ | | ✔ | | | $Q_x, Q_y$ | $G_{yz}, G_{zx}$ |
| Hexagonal | D$_6$ | | ✔ | ✓ | | $G_0$ | | $G_u$ |
| | D$_{3h}$ | | | | | | | |
| | C$_{6v}$ | ✔ | | | ✔ | | $Q_z$ | |
| | C$_6$ | ✔ | ✔ | ✔ | | $G_0$ | $Q_z$ | $G_u$ |
| | C$_{3h}$ | | | | | | | |
| Trigonal | D$_3$ | | ✔ | ✔ | | $G_0$ | | $G_u$ |
| | C$_{3v}$ | ✔ | | | ✓ | | $Q_z$ | |
| | C$_3$ | ✔ | ✔ | ✔ | | $G_0$ | $Q_z$ | $G_u$ |

Table 2. Four types of multipoles classified by parity and time-reversal properties.

| Type | Symbol | Parity $\mathcal{P}$ | | Time-Reversal $\mathcal{T}$ |
|---|---|---|---|---|
| Electric (E) Multipole | $Q_0, Q_z, Q_{xy} ...$ | Polar (True) | $(-1)^l$ | Even |
| Magnetic (M) Multipole | $M_0, M_z, M_{xy} ...$ | Axial (Pseudo) | $(-1)^{l+1}$ | Odd |
| Electric Toroidal (ET) Multipole | $G_0, G_z, G_{xy} ...$ | Axial (Pseudo) | $(-1)^{l+1}$ | Even |
| Magnetic Toroidal (MT) Multipole | $T_0, T_z, T_{xy} ...$ | Polar (True) | $(-1)^l$ | Odd |

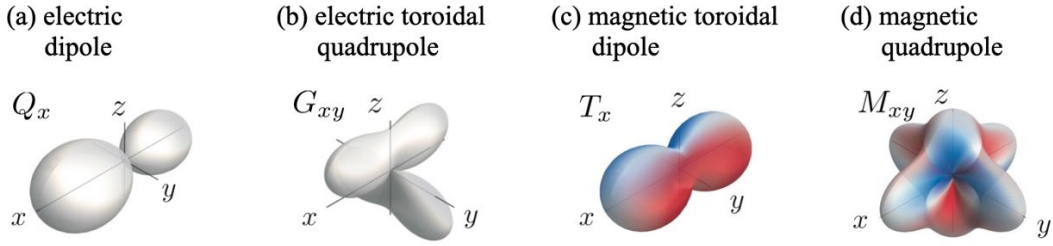

Figure 1. Examples of atomic orbitals and associated symmetry-adapted multipoles. The shape and colormap represent the density of electric charge and z-component of the orbital angular momentum of wave functions, respectively.

By using these four types of multipoles, all point groups are uniquely classified by the active multipole moments. In the language of group theory, any active moments belong to the totally symmetric (identity) irreducible representations, and they are so-called order parameters in Landau theory of phase transition. The classification of the non-centrosymmetric point groups in terms of the active multipole moments is given by Table 1. By this classification, one can recognize that active electric toroidal monopole, $G_0$, is the manifestation of the chirality. As shown in Table 2, the electric toroidal monopole, $G_0$, is $\mathcal{P}$-odd and $\mathcal{T}$-even pseudoscalar, that is the true meaning (concrete representation) of Kelvin's and Barron's definitions of chirality. Since the chirality itself is irrelevant to the time-reversal operation, its order parameter must be $\mathcal{T}$-even. The sign of the monopole $G_0$ represents two possible handedness (right or left) of the chiral objects.

Now the difference between chiral and other point groups is evident. For example, the NOA active objects are characterized either by the electric toroidal quadrupole or monopole as shown in the column SG (strong-gyrotropic) in Table 1. The chiral point groups always belong to strong gyrotropic point groups, but the opposite is not true. The nonchiral strong gyrotropic point groups, $D_{2d}$, $S_4$, $C_{2v}$, $C_s$, also exhibit NOA. The presence of NOA is just necessary but not sufficient condition for a molecule or crystal to be chiral. In this sense, the NOA should more appropriately be called natural strong-gyrotropic effect. As a similar example, so called magneto "chiral" effect [15] is an interference effect between the NOA and Faraday effect. Therefore, this effect should be called magneto strong-gyrotropic effect.

Although the explicit treatment of $G_0$ associated with specific wave functions of chiral objects depends on the molecules and crystals in question, a systematic prescription to compute the active multipole moments have been developed recently and is straightforwardly applied to any molecules and crystals. A typical example of the elemental Tellurium is found in ref. [16]. It is notable that $G_0$ term in chiral Tellurium is quite large with respect to other terms in the Hamiltonian. The energy scale estimated by first-principle method is about 1.7 eV, which well exceeds the spin-orbit coupling of Tellurium atom. Such a large energy enhancement seems to be obtained as a consequence of chiral collective effect spanning over the whole wave functions. Another example for a fictitious chiral electron-phonon system will be shown later.

It should be also pointed out that although the electric toroidal monopole, $G_0$, seems to be featureless, it has indeed an internal structure. This observation makes us extend the concept of chirality to various states of matters. For example, the helicity $\boldsymbol{k} \cdot \boldsymbol{\sigma}$ has propagating momentum $\boldsymbol{k}$ and its spin angular momentum $\boldsymbol{\sigma}$ along $\boldsymbol{k}$. Since $\boldsymbol{k}$ is $\mathcal{T}$-odd polar vector, and $\boldsymbol{\sigma}$ is $\mathcal{T}$-odd axial vector, their scalar product is regarded as $G_0$. Another example is the monoaxial chiral helimagnetic order along $z$ axis. The energy gain of the Dzyaloshinskii-Moriya (DM) interaction due to this ordering is given by $\boldsymbol{D} \cdot (\boldsymbol{S} \times \nabla_z \boldsymbol{S})$. Since $\boldsymbol{D}$ is $\mathcal{T}$-even axial DM vector, and $(\boldsymbol{S} \times \nabla_z \boldsymbol{S})$ is $\mathcal{T}$-even polar vector, their product is also regarded as $G_0$. In general, $G_0$ can be decomposed into $\boldsymbol{Q}_a \cdot (\boldsymbol{Q}_b \times \boldsymbol{Q}_c)$, $\boldsymbol{Q} \cdot$

($M_a \times M_b$), and so on where $Q_{a,b,c}$ and $M_{a,b}$ are the electric and magnetic dipoles, respectively. It is interesting to see, in a physical point of view, that such a structure indeed appears both in the Hamiltonian of Tellurium as a lattice-spin coupling, $R \cdot (l \times \sigma)$ [16], and in the parity-violation term due to weak interaction of elementary particles (see next section), which is a source of chiral response.

## 3. Recognition of chirality in enantioselective fields

Once the order parameter of chiral objects is identified, one can discuss its response to the external fields through the invariant coupling of the electric toroidal monopoles between objects and fields. Namely, there is an invariant coupling in the Hamiltonian with the form of

$$H_{\text{int}} = g G_0(\text{field})\, G_0(\text{object})$$

where $G_0(\text{object})$ and $G_0(\text{field})$ represent the chiral order parameter of the object and its conjugate external field with the coupling strength, $g$, respectively. From the energetics point of view, the external field $gG_0(\text{field})$ with negative sign favors the positive sign of $G_0(\text{object})$, i.e., one handedness, while the opposite external field favors another handedness. Thus, the external field, $gG_0(\text{field})$ is regarded as enantioselective field. It is important to note that the enantioselective fields can be made up from plural external sources as long as the symmetry of the combined fields coincides with that of the electric toroidal monopole, i.e., the $\mathcal{T}$-even pseudoscalar in the relevant point group as was discussed in the last part of the previous section. We will discuss several examples below. Note that any external fields without $G_0(\text{field})$ are orthogonal to $G_0(\text{object})$, so that no interaction is expected.

One example of $gG_0(\text{field})$ is electromagnetic field called Lipkin's Zilch $\rho_\chi$, which is a signature of circularly polarized light. It is defined as

$$\rho_\chi = \frac{\varepsilon_0}{2} E \cdot (\nabla \times E) + \frac{1}{2\mu_0} B \cdot (\nabla \times B)$$

where $E$ and $B$ are electric and magnetic fields of electromagnetic wave. Tang and Cohen have clearly shown that Zilch is proportional to the circular dichroism [17]. We summarize typical external fields classified by multipoles in Table 3. Using this table, one can confirm that the combinations of $E \cdot (\nabla \times E)$ and $B \cdot (\nabla \times B)$ have the same symmetry of $G_0$. Electrical magneto-chiral anisotropy (EMChA) is also understood as a conjugate $G_0$ field. In other words, the scalar product of the electric current $J$ and magnetic field $B$ represents $gG_0(\text{field})$ [18].

Table 3. Representative external fields classified by the multipole basis. $u_i$ denotes the $i$-th component of the lattice displacement vector.

| External Field | Symbol | Multipole | Rank | Parity $\mathcal{P}$ | Time-Reversal $\mathcal{T}$ |
|---|---|---|---|---|---|
| Zilch | $\rho_\chi$ | $G_0$ | Monopole | Odd | Even |
| Electric Field | $E$ | $Q$ | Dipole | Odd | Even |
| Magnetic Field | $B$ | $M$ | Dipole | Even | Odd |
| Rotational Field | $\omega = (\nabla \times u)/2, \nabla \times E$ | $G$ | Dipole | Even | Even |
| Electric Current | $J, \nabla \times B$ | $T$ | Dipole | Odd | Odd |
| Strain | $\varepsilon_{ij} = (\nabla_i u_j + \nabla_j u_i)/2$ | $Q_{ij}$ | Quadrupole | Even | Even |

We here mention another example of chiral coupling. Parity violation of the electroweak force leads to the parity violation energy difference (PVED). The coupling between subnuclear quarks and molecular electrons is generally written in the form [19]

$$E_{\rm PV} = 2{\rm Re}\sum_j \frac{\langle 0|H_{\rm PV}|j\rangle}{E_0 - E_j}\langle j|H_{\rm SO}|0\rangle$$

where $H_{\rm PV} \propto \{\boldsymbol{\sigma}\cdot\boldsymbol{p}, \delta(\boldsymbol{r})\}$ is the parity-violated interaction Hamiltonian and $H_{\rm SO} \propto \boldsymbol{\sigma}\cdot\boldsymbol{l}$ is an electron spin-orbit coupling Hamiltonian. In this case, the quark-electron interaction plays a role of $gG_0({\rm field})$ for the electron subsystem.

It is important to notice that multipoles are mutually interchangeable, namely, the rotation operation ($\boldsymbol{\nabla}\times$) converts their parity, while the time integral/derivative alters their time-reversal property. For example, the electric current $\boldsymbol{J}$ is $\mathcal{T}$-odd, while accumulated charge as a result of current generates electric field $\boldsymbol{E}$ ($\mathcal{T}$-even). In this case, the time integral plays a role in converting magnetic toroidal dipole, $\boldsymbol{T}$, into electric one, $\boldsymbol{Q}$ according to Table 3 and Fig. 2.

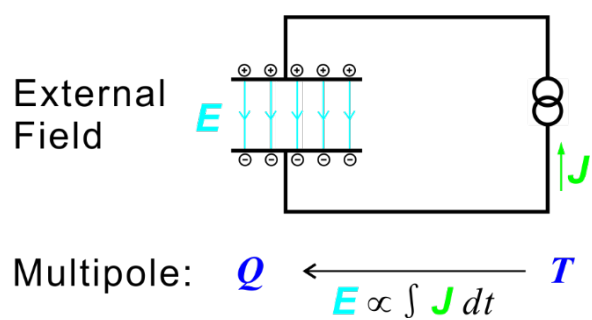

Fig. 2 An example of multipole conversion from $\boldsymbol{T}$ to $\boldsymbol{Q}$.

Barron has discussed the connectivity between chirality and enantioselective field by proposing "true chirality" as $\mathcal{T}$-even $\mathcal{P}$-odd pseudoscalar [9-11]. In his papers, he pointed out that, in addition to the structural chirality, another type of enantiomorphism which he calls "false chiral influence" exists. The important difference between "true" and "false" chirality is their response to time reversal operation where false chiral influence is $\mathcal{T}$-odd $\mathcal{P}$-odd. In a multipole's way of expression, the true chirality corresponds to $G_0$ and the false chirality is $M_0$. One example of this falsely chiral field is a parallel combination of electric field $\boldsymbol{E}$ and magnetic field $\boldsymbol{B}$. In the multipole classification, $\boldsymbol{E}\cdot\boldsymbol{B}$ is a magnetic monopole $M_0$ according to Table 3.

He stressed an importance of this discrimination because absolute asymmetric synthesis is impossible with "falsely chiral" field at equilibrium of liquids (which corresponds to SO(3) in which any proper rotation is allowed). This is because the falsely chiral field cannot recognize, or cannot lift the degeneracy of, any chiral enantiomers in it. This is consistent with our understanding because $M_0({\rm field})$ cannot have any coupling with $G_0({\rm object})$ because of different symmetry in time domain. But at the same time, he discussed a possibility of absolute asymmetric synthesis from falsely chiral field with kinetic condition or inhomogeneous media. This possibility of absolute asymmetric synthesis by external field can be understood, in our point of view, by creating truly chiral field $G_0({\rm field})$ by applying time differential or time integration to a given falsely chiral field $M_0({\rm field})$. In real systems, sometimes $\mathcal{T}$-even and $\mathcal{T}$-odd components are mixed when the objects and/or fields are moving. Therefore, one needs to decompose the total response into $G_0({\rm object})$ and/or $G_0({\rm field})$ and other species to understand the essential role of chirality in a given phenomenon.

Let us now discuss chirality recognition by CISS effect, which is an important extension of CISS to chemical application. Recently, Naaman and coworkers have reported separation of chiral molecules by magnetic substrate which is magnetized perpendicular to the surface [20, 21]. Interestingly, this effect is kinetic and disappears at long time experiment, which implies an influence of kinetic processes. At present, how to explain this enantio-separation at ferromagnet surface is under debate. The authors will clarify symmetry aspects of this phenomenon in accordance with the multipole method.

In order to recognize chirality of the system at finite temperature, or in order to differentiate one enantiomer from the other, one needs to identify $G_0(\text{object})$ both in ground and excited states (including those materials such as self-assembled monolayer and solid state in case of CISS). In addition, the time integral or time derivative should be applied to the quantity of interest, if one thinks about dynamics (excited state or non-equilibrium state) of a given chiral system. In this particular case, it seems that $G_0$ term in spin current (helicity of electron) can be transformed into metastable $M_z$ when spin reservoirs at molecular edges are assumed. Naaman has proposed antiparallel spin pair accumulated at two opposite ends of chiral molecule as shown in Fig 3, when the molecule is fluctuating around the substrate surface [20]. In such a case the metastable $M_z$ at lower end of the molecule should interact with $M_z$ at the ferromagnet surface via the exchange interaction, $J$, and therefore enantio-selection becomes possible. In other words, $dE_z/dt$ excites the molecule to induce displacement current which is spin polarized due to CISS.

Here, the authors point out that this antiparallel spin pair is composed of two magnetic dipoles $M_z$ and $-M_z$ which constitutes $\mathcal{T}$-odd monopole, $M_0$ as a whole. Therefore, the connection between molecular chirality $G_0$ and the $\mathcal{T}$-odd monopole, $M_0$ should be clarified. In this discussion, the authors recall attention to the relationship between $\mathcal{T}$-odd and $\mathcal{T}$-even quantities. For example, according to Table 3 and Fig 2, electric current is represented by $T_z$ ($\mathcal{T}$-odd), while accumulated charge as a result of current generates electric field $Q_z$ ($\mathcal{T}$-even). A time integral plays a role in converting $\mathcal{T}$-property of multipoles, in the present case, from $G_0$ to $M_0$. As such, spin polarized current with finite helicity ($\mathcal{T}$-even $G_0$) will end up with spin accumulation ($\mathcal{T}$-odd $M_z$) or the $\mathcal{T}$-odd monopole, $M_0$ as a whole, when edges of an object can act as spin reservoirs.

Therefore, given that CISS effect exists in chiral molecules and that the molecule is moving around the surface of an ferromagnet, where **E** perpendicular to the surface exists because of contact potential (work function difference), the internal displacement current due to dielectric response will accumulate antiparallel spin polarization at two opposite ends of the molecule. Most importantly, inward/outward arrangement of this spin accumulation is governed by the handedness of the molecule because of the spin-momentum locking in chiral system. Therefore, either parallel (repulsive *J*) or anti-parallel (attractive –*J*) interaction is applied between chiral molecule and the substrate, depending on the handedness of the molecule as well as the magnetization direction of the substrate. In reality, all of **E**, **B**, *dE*/*dt*, and *dB*/*dt* exist in the present situation and any combination of those fields may come into play. But one is able to extract the interaction associated only to the chirality in the form of either $gG_0(\text{field}) \cdot G_0(\text{object})$ or $JM_z(\text{field}) \cdot M_z(\text{object})$ forms based on the discussion above. In the former case, the chiral molecule's $G_0$ interacts with a chiral field ($\mathcal{T}$-even $G_0(\text{field})$) that is generated by the combination of ferromagnetic spins ($\mathcal{T}$-odd $\pm M_z$) and $dE_z/dt$ ($\mathcal{T}$-odd). In the latter case of interaction, magnetic exchange between the magnetic moment $M_z$ of ferromagnetic substrate interacts with accumulated spin $M_z$ at the edge of the chiral molecule. As a result, in both types of interpretations, this chirality-specific interaction will result in a chirality recognition and thus the enantio-separations reported in recent papers become possible.

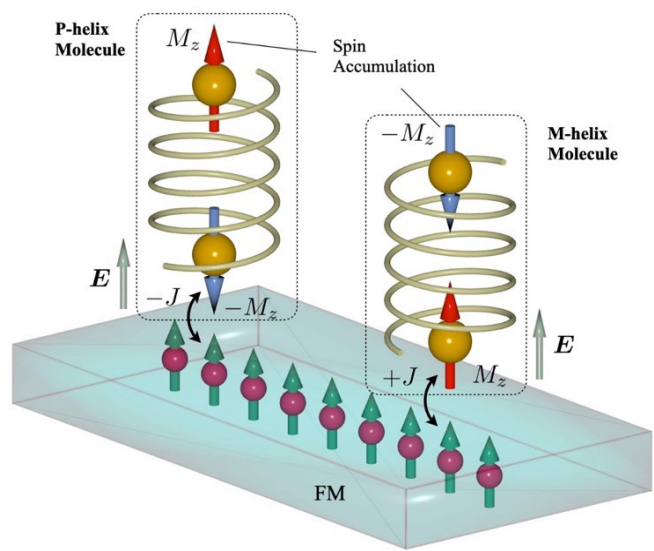

Fig. 3 Helical molecules approaching to a ferromagnetic substrate (FM). Because of the displacement current associated with CISS effect, metastable spin accumulation, $M_z$ or $-M_z$, at both ends of the molecules is expected.

4. **Dynamics of chiral order parameter**

In this section, the authors discuss dynamics of $G_0$. The aim of this discussion is to identify the influence of chirality fluctuations. Another objective is to recall an attention to the stability of chiral and "enantiopure" states in terms of symmetry breaking. Both topics are coming from the tendency of nature to restore the original higher symmetry even after the symmetry breaking.

Because the discovery of chirality was made with stable chemical species such as quartz and tartrate, it tends to be considered as a firm entity once the structure of enantiomer is established. However, if the energy barrier that separates the two enantiomers is comparable to or smaller than the environmental thermal energy, the enantiomers start to interconvert one another, which is called racemization. In this regard, an emergence of chirality, or $G_0$, is similar to phase transition where lowering the temperature results in symmetry breaking, although it is not a collective phenomenon. It is important to point out that $G_0$ of chiral molecules and crystals is protected by double-well potential after symmetry breaking. In this sense, chirality can be redefined as $G_0$ protected by energy barrier. Despite this protection, the tendency to the enantiomeric interconversion remains still active even with higher energy barrier when quantum tunneling and/or thermal excitation are considered. In the physics of phase transition, such a tendency is known to work as a fluctuation that try to recover higher symmetry even after the symmetry breaking. Such thermal/quantum fluctuations are sometimes a source of enhanced interactions and may create emergent properties such as superconductivity [22-25]. From this point of view, the tendency to restore the original higher symmetry may also have an effect on physical and/or chemical properties of enantiomers, too, if the dynamics of the chemical structure is considered. Systematic elaboration of this effect is, however, not yet discussed to the best of our knowledge.

Now the authors discuss flexible dynamics of chiral structure in more detail. Morino and Mizushima found that free rotation of 1,2-Dichloroethane stops at sufficiently low temperature to find rotational isomers called anti and gauche [26]. Although they didn't point out it, there are two chiral enantiomers in the gauche form as shown in Figure 4. This means that two enantiomers can interconvert *as a function of time* when the temperature is high enough to overcome the energy barrier that separates two states. [5]Helicene is another example of enantiomeric interconversion where the energy barrier is comparable to $k_BT$ at room temperature. In this regard, even asymmetric $sp^3$ carbon in fluoro-chloro-bromo-methane should also racemize at sufficiently high temperatures. (Fig. 4c)

In a quantum mechanical point of view, a tunneling effect can also take place in the racemization process. Let us consider chiral ammonia NH(D)(T) molecule that can flip between R (right) and L (left) forms. Even at zero temperature, the quantum levels for R and L enantiomers can have resonant tunneling because their energy levels are always identical due to the symmetry of Schrödinger equation. Neither R form nor L form can be an eigen state of the system alone because of symmetry consideration [9], and is therefore the complete wave function should be constructed from their linear combination. In that case, two enantiomeric states can oscillate one another as shown in Fig 5b (right panel). Such an oscillation can be described by

$$\varphi(t) = \frac{1}{\sqrt{2}}[\varphi^0(+) + \varphi^0(-)e^{-i\omega t}]e^{-iW(+)t/\hbar}$$

where $\varphi^0(+)$ and $\varphi^0(-)$, are wave functions with even and odd parities, which are the eigen states of the Hamiltonian. $\hbar\omega$ is the energy difference of these two states, while $W(+)$ is the energy for even parity state [9]. In reality, the life time of an enantiomer due to the thermal/quantum interconversion is determined by the energy barrier height and system size.

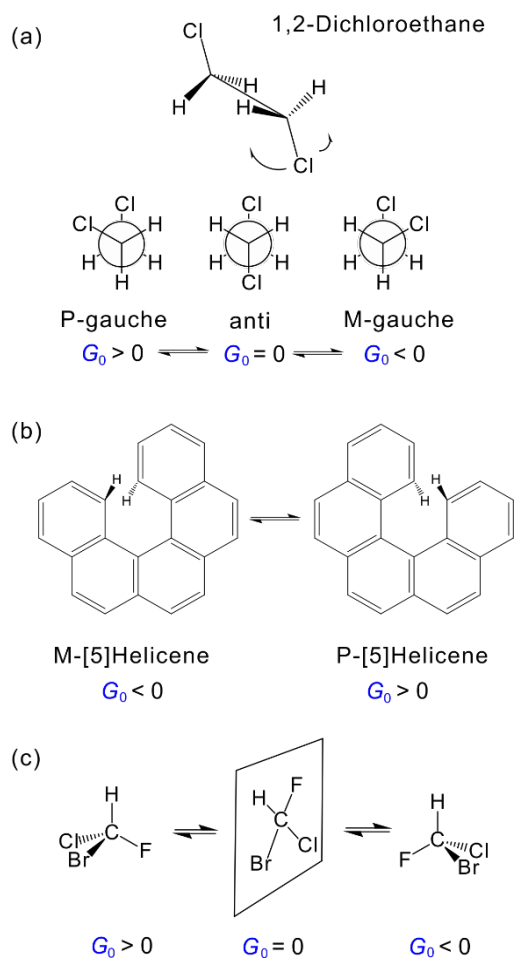

Figure 4. Thermal racemization process for (a) 1,2-Dichloroethane (expressed by sawhorse projection and Newman projection), (b) [5]Helicene, and (c) fluoro-chloro-bromo-methane. The sign of $G_0$ is arbitrary.

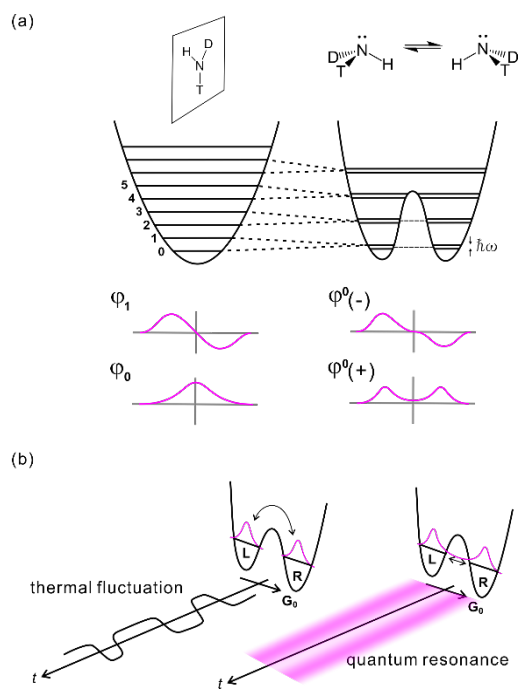

Figure 5. (a) Racemization process by quantum tunneling. (b) $G_0$ oscillation by thermal (left) and quantum (right) racemization.

When the above life time of enantiomer is considered to be finite, one can also discuss a life time of enantio pure state. If a group of chiral molecules which are oscillating between two enantiomers is allowed to propagate in time, there are some limited periods of time when all the handedness of molecules is the same (Fig 6a). This is what we call enantio pure state, although this period can be as long as a life of universe in case the energy barrier is sufficiently high. This is quite analogical to the phase transition of ferromagnetism. In a ferromagnet, a phase transition at Curie temperature makes the spins within a domain aligned in the same direction. After poling process by external magnetic field ($M_z$), a single-domain magnet with parallel magnetization can be created. This is a time-reversal-symmetry (TRS)-broken state because one can flip the spin moment by time reversal operation $\mathcal{T}$. On the other hand, a molecule can exhibit stable chiral structure, when the system temperature becomes well below the energy barrier separating two enantiomers, to yield a racemate. After an enantio-separation process for this racemate by external operation such as chiral chromatography ($G_0$), an enantio-pure state can be obtained. Note that enantio-pure state is $\mathcal{T}$-even while ferromagnetic single-domain state is $\mathcal{T}$-odd. The difference between these two examples is originating in the nature of order parameters ($M_z$ and $G_0$; Fig 6b) for magnetic and chiral symmetry breaking. In the vicinity of magnetic ordering, the fluctuation is known to induce superconductivity in some cases [22-25]. In a similar way, one may envisage an emergent effect from the chirality fluctuation which try to retrieve the structure of the opposite enantiopure state, when a dynamics of chiral molecule is considered. When the molecule is vibrating, for example, the symmetry of the opposite enantiomer should be mixing as a fluctuation to yield a certain type of correction [27].

Let us now discuss the effect of the chirality dynamics. Here we define 'dynamic' as a time-dependent variance of materials and fields *within the enantiopure time domain* described above. In this definition, the chirality should be retained the same throughout the dynamics while its magnitude (order parameter) is changing. This chirality fluctuation may have other types of effect than the static chirality. Such a fluctuation should be visible when a chiral molecule is vibrating, for example.

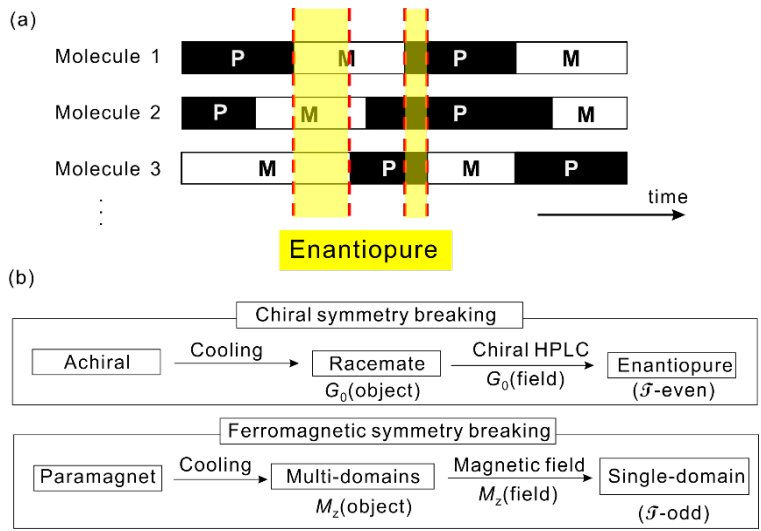

Figure 6. (a) Enantiopure state as a time domain. (b) Comparison between the chiral symmetry breaking and magnetic symmetry breaking. HPLC = High Pressure Liquid Chromatography

Kato, JK and HMY have introduced a pseudo Jahn Teller Hamiltonian to calculate the dynamics of $G_0$ in a C$_3$ chiral molecule that is being passed through by a tunneling electron [28]. The pseudo Jahn-Teller Hamiltonian $H_{en}$ and spin-orbit Hamiltonian $H_{SOC}$ are as follows.

$$H_{en} = V_+ Q_- |\phi_z\rangle\langle\phi_+| + V_- Q_+ |\phi_z\rangle\langle\phi_-| + V_-^* Q_- |\overline{\phi_z}\rangle\langle\phi_+| + V_+^* Q_+ |\overline{\phi_z}\rangle\langle\phi_-| + V_0 Q_- |\phi_+\rangle\langle\phi_-| + H.c.$$

$$H_{SOC} = \lambda(|\phi_+\rangle\langle\phi_+| - |\phi_-\rangle\langle\phi_-|) \otimes \hat{\sigma}_z$$

where $Q_+$, $Q_-$, $|\phi_z\rangle$, $|\phi_+\rangle$, $|\phi_-\rangle$, and $\hat{\sigma}_z$ are bases for nuclear clockwise rotation, nuclear counterclockwise rotation, electronic z-translation, electronic clockwise rotation, electronic counterclockwise rotation, and electronic spin respectively (refer to Fig. 7(a)). $V_+$, $V_-$, $V_0$, and $\lambda$ are parameters for each energy strength. The chiral symmetry breaking manifests for $V_+ \neq V_-$. Although the above Hamiltonian consists of two parts, the electron-nuclear coupling and the electron spin-orbit coupling terms, it is possible to obtain an effective Hamiltonian after integrating out the degrees of freedom associated with the nuclear vibration coordinates. A resultant effective Hamiltonian consists of $gG_0$(field) containing nuclear vibrations, while $G_0$(object) contains only electron degrees of freedom.

Diagonalization of this $G_0$ term leads to four eigenstates labeled with an injected electron momentum direction (P) and total spin (J). The total spin J consists of electron spin and nuclear vibrational orbital angular momentum originating from discrete rotational symmetry. This result was actually demonstrated by fully diagonalizing the nuclear vibration. The splitting $\delta - \lambda$, where $\delta$ is an energy difference between orbital angular momenta j = 0 and 1 states, directly corresponds to $gG_0$(field), and gives spin dependent barrier potential. This energy split dominates the total energy scale and spin orbit-coupling $\lambda$ plays only a minor role in separating two Kramer's paired states in momentum space (both of which possess the same helicity). Thus, the spin barrier is proportional to $gG_0$(field) (Fig. 7(b)). Consequently, there arises a spin-dependent barrier potential for an injecting electron and an electron with favored (disfavored) spin tends to go through (repelled from) a chiral molecule (Fig. 7(c)), which can explain the large spin polarization in CISS effect.

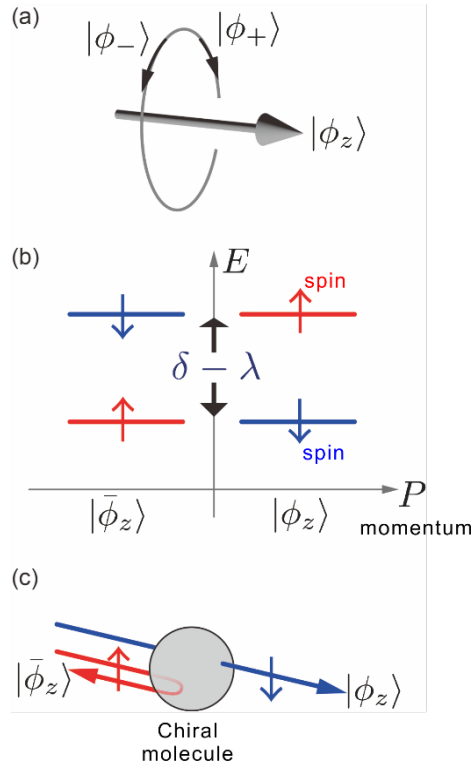

Figure 7. Pseudo JT model to explain CISS effect. For details of the parameters, refer to [28].

In the theoretical investigation of CISS effect, one needs to explain the fact that the magnitude of spin polarization is much larger than one can expect from simple spin-orbit Hamiltonian. In addition, Bardarson theorem prohibits any spin polarization as far as a scattering process by two-level two-terminal system is considered [29]. A source of third terminal is, therefore, being looked for by several theoretical researchers. In order to overcome these problems, recent experiments and calculations suggest an involvement of phonons as a source of effective spin polarization [30-33] whereas others will try to include exchange interactions [34, 35]. A model proposed here is another way to address this issue by explicitly including pseudo Jahn Teller effect into the system to include dynamics of chiral order parameter $G_0$(object). As an extension of this idea to crystalline materials, we are also interested in the effect generated by chiral phonons [36], which can be extended to CISS effects in solids [37, 38].

## 5. Summary

In summary, we have demonstrated that electric toroidal monopole $G_0$ is the order parameter for chiral symmetry breaking. Because the chiral structure is strongly connected with chemical bonding, the interplay between physics (symmetry breaking) and chemistry (bonding) appears in an entangled manner in real systems. Even in such a case, factor decomposition by multipole concept should be a firm guideline to find out the essential role of chirality in a given situation. At present, methodology to analyze non-equilibrium and/or excited states is still quite limited, but development of thermodynamics to describe those states is needed to understand the recent experiments. For example, it has been experimentally shown that a combination of gravity and rotation can generate chiral aggregation from achiral molecules, in which the handedness can be reversed by rotation reversal [39, 40]. The way to explain the mechanism of these emergent chirality will be a big and fruitful future challenge.

**Acknowledgements**

This work was supported by JSPS KAKENHI Grants No. 19H00891, 21H01032, and No. 21H01034 and Cooperative Research by Institute for Molecular Science (IMS program 22IMS1220).

References

[1] B. Göhler, V. Hamelbeck, T. Z. Markus, M. Kettner, G. F. Hanne, Z. Vager, R. Naaman, H. Zacharias, *Science*, **2011**, *331*, 894.

[2] K. Michaelia, D. N. Beratan, D. H. Waldeck, and Ron Naaman, *Proc. Nat. Acad. Sci*. **2019**, *116*, 5931.

[3] O. Ben Dor, S. Yochelis, A. Radko, K. Vankayala, E. Capua, A. Capua, S. -H. Yang, L. T. Baczewski, S. S. P. Parkin, R. Naaman, and Y. Paltiel, *Nature Commun*. **2017**, *8*, 14567.

[4] R. Naaman, Y. Paltiel, and D. H. Waldeck, *Acc. Chem. Res*. **2020**, *53*, 2659

[5] M. Suda, Y. Thathong, V. Promarak, H. Kojima, M. Nakamura, T. Shiraogawa, M. Ehara, and H. M. Yamamoto, *Nature Commun*. **2019**, *10*, 2455.

[6] Lord Kelvin, *The molecular tactics of a crystal*, (Oxford, Clarendon Press, **1894**).

[7] F. Hund, *Z. Phys*. **1927**, *43*, 805

[8] L. Rosenfeld, *Z. Phys*. **1928**, 52,161.

[9] L. D. Barron, *Chem. Soc. Rev*. **1986**, 15, 189

[10] L.D. Barron, *J. Am. Chem. Soc*., **1986**, 108, 18, 5539.

[11] L.D. Barron, Molecular light scattering and optical activity, 2$^{nd}$ Ed, (Cambridge University Press, **2004**)

[12] S. Hayami and H. Kusunose, *J. Phys. Soc. Jpn*. **2018**, 87, 033709.

[13] S. Hayami, M. Yatsushiro, Y. Yanagi, and H. Kusunose, *Phys. Rev. B* **2018**, *98*, 165110.

[14] H. Kusunose, R. Oiwa, and S. Hayami, *J. Phys. Soc. Jpn*. **2020**, *89*, 104704.

[15] L. D. Barron and J.Vrbancich, *Mol. Phys*. **1984**, *51*, 715

[16] R. Oiwa and H. Kusunose, arXiv:2203.15192.

[17] Y. Tang and A. E. Cohen. *Phys. Rev. Lett*. **2010**, *104*, 163901.

[18] G. L. J. A. Rikken, J. Fölling, and P. Wyder, *Phys. Rev. Lett*. **2001**, *87*, 236602.

[19] R.A. Harris and L. Stodolsky, *J. Chem. Phys*. **1980**, *73*, 3862.

[20] K. Banerjee-Ghosh1, O. Ben Dor, F. Tassinari, E. Capua, S. Yochelis, A. Capua, S. -H. Yang, S. S. P. Parkin, S. Sarkar, L. Kronik, L. T. Baczewski, R. Naaman, Y. Paltiel, *Science* **2018**, *360*, 1331

[21] D. Bhowmick, Y. Sang, K. Santra, M. Halbauer, E. Capua, Y. Paltiel, R. Naaman, and F. Tassinari, *Cryst. Growth Des*. **2021**, *21*, 2925.

[22] N. F. Berk and J. R. Schrieffer, *Phys. Rev. Lett*. **1966**, *17*, 433.

[23] S. Nakajima, *Prog. Theor. Phys*., **1973**, *50*, 1101

[24] K. Miyake, S. Schmitt-Rink, and C. M. Varma, *Phys. Rev. B* **1986**, *34*, 6554(R).

[25] D. J. Scalapino, E. Loh, Jr., and J. E. Hirsch, Phys. Rev. B **1986**, *34*, 8190(R).

[26] S. Mizushima and Y. Morino, *Bull Chem. Soc. Jpn*. **1941**, *17*, 94.

[27] This may be an effect similar to the spin-orbit interaction which is a correction for charge symmetry $C$ (electron/positron symmetry) breaking.

[28] A. Kato, H. M. Yamamoto, and J. Kishine, *Phys. Rev. B* **2022**, *105*, 195117.

[29] J. H. Bardarson, *J. Phys. A: Math. Theor*. **2008**, *41*, 405203.

[30] N. Sasao, H. Okada, Y. Utsumi, O. Entin-Wohlman, and A. Aharony, *J. Phys. Soc. Jpn*. **2019**, *88*, 064702.

[31] G. -F. Du, H. -H. Fu, and R. Wu, *Phys. Rev. B* **2020**, *102*, 035431.

[32] L. Zhang, Y. Hao, W. Qin, S. Xie, and F. Qu, *Phys. Rev. B* **2020**, *102*, 214303

[33] J. Fransson, *Phys. Rev. B* **2020**, *102*, 235416.

[34] J. Fransson, *J. Phys. Chem. Lett.* **2019**, *10*, 7126.

[35] M. S. Zöllner, A. Saghatchi, V. Mujica, and C. Herrmann, *J. Chem. Theory Comput.* **2020**, *16*, 7357.

[36] J. Kishine, A. S. Ovchinnikov, and A. A. Tereshchenko, *Phys. Rev. Lett*. **2020**, *125*, 245302.

[37] A. Inui, R. Aoki, Y. Nishiue, K. Shiota, Y. Kousaka, H. Shishido, D. Hirobe, M. Suda, J. Ohe, J. Kishine, H. M. Yamamoto, and Y. Togawa, *Phys. Rev. Lett*. **2020**, *124*, 166602.

[38] Y. Nabei, D. Hirobe, Y. Shimamoto, K. Shiota, A. Inui, Y. Kousaka, Y. Togawa, and H. M. Yamamoto, *Appl. Phys. Lett*. **2020**, *117*, 052408.

[39] M. Kuroha, S. Nambu, S. Hattori, Y. Kitagawa, K. Niimura, Y. Mizuno, F. Hamba, and K. Ishii, *Angew. Chem. Int. Ed*. **2019**, *58*, 18454.

[40] N. Micali, H. Engelkamp, P. G. van Rhee, P. C. M. Christianen, L. Monsu Scolaro and J. C. Maan, *Nature Chem*. **2012**, *4*, 201.